\documentclass[pre,floatfix,superscriptaddress,twocolumn,9pt]{revtex4-1}

\usepackage{amsmath}
\usepackage{amssymb}
\usepackage{amsfonts}
\usepackage{graphicx}
\usepackage{color}
\usepackage{hyperref}
\usepackage{bm}
\usepackage[english]{babel}
\usepackage{siunitx}

\newcommand{\uvec}[1]{\ensuremath{\hat{\mathbf{#1}}}}

\newcommand{\Tr}{\ensuremath{\rm Tr}}

\newcommand{\figurewidth}{.45\textwidth}

\newcommand{\nn}{\ensuremath{\hat{\mathbf{n}}}}

\newcommand{\cross}{\ensuremath{\times}}
\newcommand{\x}{\ensuremath{\cross}}
\newcommand{\splay}{$k_{11}$}
\newcommand{\twist}{$k_{22}$}
\newcommand{\bend}{$k_{33}$}

\newcommand{\thetitle}{Automated Determination of $n$-Cyanobiphenyl Elastic Constants in the Nematic Phase from Molecular Simulation}

\begin{document}

\title{\thetitle}
\author{Hythem Sidky}
\altaffiliation{Current address: Institute for Molecular Engineering, University of Chicago, Chicago IL, 60637.}
\affiliation{%
	Department of Chemical and Biomolecular Engineering, %
	University of Notre Dame, %
	Notre Dame, IN 46556
}
\author{Jonathan K. Whitmer}
\altaffiliation{Author to whom correspondence should be addressed: jwhitme1@nd.edu}
\affiliation{%
	Department of Chemical and Biomolecular Engineering, %
	University of Notre Dame, %
	Notre Dame, IN 46556
	}

\begin{abstract}
New applications of liquid crystalline materials have increased the need for
precise engineering of elastic properties. Recently, Sidky et al. presented
methods by which the elastic coefficients of molecular models with atomistic
detail can be accurately calculated, demonstrating the result for the
ubiquitous mesogen 5CB. In this work, these techniques are applied to the
homologous series of $n$CB materials, focusing on the standard bend, twist,
and splay deformations, using an entirely automated process. Our results show
strong agreement with published experimental measurements for the $n$CBs and
present a path forward to computational molecular engineering of liquid
crystal elasticity for novel molecules and mixtures.
\end{abstract}

\maketitle

\section*{Introduction}

Liquid crystals (LC) are a  class of fluids which exhibit long-range
orientational ordering and a resistance to spontaneous
deformation.~\cite{SoftMatterPhys} While traditionally used in display
technologies~\cite{Gray1999}, advances in both our understanding of elasticity
and material fabrication has opened up many new and interesting applications
where the unique properties of LCs can be exploited. Chemoresponsive LCs have
been designed to respond to targeted chemical species, which can be quantified
through polarized optical microscopy.~\cite{tibor2017} LC elastomeric materials
can function as elements in soft robots and sensors.~\cite{Babakhanova2018}
Topological defects and inclusion formation in LCs have also been used as
templates for molecular self-assembly.~\cite{Wang2015} Many of these novel
applications rely on the interplay between external, boundary and elastic restoring
forces, which is dictated by both the absolute magnitudes and ratios of the
elastic constants. 

Advances in atomistic simulations have made it possible to gain an unprecedented
level of detail into the molecular underpinnings of liquid crystalline behavior
which can be rigorously verified against experiment.~\cite{Palermo2013,
Ramezani-Dakhel2017} By understanding the role of molecular features such as
conformational flexibility and charge distribution, new LCs can be engineered
{\it in silico} to have highly specific thermophysical properties necessary for
biosensing, templating, and other emerging technologies. Until
very recently~\cite{Sidky2018}, it was not possible to economically predict the
elastic properties of molecular LCs and thus made it extremely difficult to
engineer novel LCs for these applications. 

The work of Sidky, et al.~\cite{Sidky2018} was a crucial first step in this process, as there it was demonstrated how molecular simulations may be used as a tool to predict the
bulk elastic moduli of the molecular liquid crystal models, focusing on the ubiquitous 4-pentyl-4'-cyanobiphenyl,
more commonly known as 5CB. There, it was also shown that the saddle-splay elastic constant $k_{24}$, difficult to obtain in experiments, could be directly measured, and this revealed the underlying positive definite nature of the mode in contradiction to recent experimental suggestions
based on indirect observation.~\cite{Allender1991, Sparavigna1994, Polak1994, Pairam2013,
Davidson2015} The stability of 5CB and its homologues, in
addition to their achromic appearance and strong positive dielectric anisotropy, has
made them widely used in research~\cite{Wang2016b, Eimura2016} and industrial
applications.~\cite{Gray1999,SoftMatterPhys} Importantly, their elastic
˘properties have been well-characterized through
experiment.~\cite{Madhusudana1982, Hakemi1983, Chen1989a, Chatopadhayay1994}
Molecular models have also been parameterized to reproduce basic thermodynamic
properties such as density, orientational order, and phase
transitions.~\cite{Tiberio2009, Boyd2015, Cacelli2007, Cacelli2009}

\begin{figure}
	\begin{center}
	\includegraphics[width=\figurewidth]{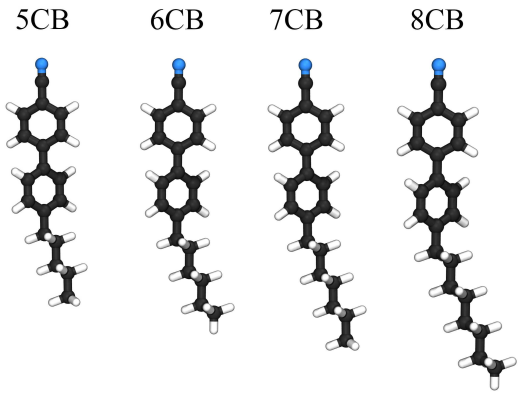}
	\end{center}
	\caption{ The nematic liquid crystal 5CB is related to many other common mesogens, including the homologous series generated by extending the alkyl tail. In this
		work, we directly measure the elastic constants of 6CB, 7CB, and 8CB
		from molecular simulation.
	}
	\label{fig:homos}
\end{figure}

The focus of this work is on applying the methodology proposed in
Ref.~\onlinecite{Sidky2018} to the $n$-cyanobiphenyls ($n$CBs) 6CB, 7CB, and
8CB, which are illustrated in Figure~\ref{fig:homos}. Although there are minimal
structural differences between this series of molecules, the extended alkyl
chain has a strong effect on ordering, which gives rise to the ``odd-even''
effect as the homologous series is ascended.~\cite{Madhusudana1982, Hakemi1983}
This phenomenon, where the molecular shape anisotropy and corresponding nematic
transition temperature increases only when $n$ goes from even to odd, has been observed in both experiments and simulations.\cite{Tiberio2009, Gray1971} While the number $n$ is arbitrary, for $n < 5$ a nematic phase is not observed, while for
$n > 7$, the LCs admits additional smectic behavior. Here, we seek to
predict the splay, twist, and bend elastic moduli of $n$CB nematogens through an automated version of the prior algorithms~\cite{Sidky2018} in order to test the limits and robustness of our methods. This represents a step towards
the the ultimate goal of computationally-guided LC engineering, and extends the 
characterization of the elastic properties of existing LC forcefields. We
restrict ourselves to the three bulk elastic constants because they are directly
measurable in experiment and their behavior is generally
well-understood. 

\section*{Methods}

The computational methods used here follow Ref.~\onlinecite{Sidky2018} closely.
For completeness, we will provide a brief summary and highlight the differences.
The linear-order elastic free energy density of a uniaxial nematic, absent of
any molecular polarity or chirality, may be represented in the Frank-Oseen
form~\cite{Stewart2004}

\begin{equation}
	\begin{aligned}
		f = \frac{1}{2} k_{11} \left(\nabla \cdot \nn \right)^2 & + %
		\frac{1}{2} k_{22} \left(\nn \cdot \nabla \x \nn \right)^2 + %
		\frac{1}{2} k_{33} \left(\nn \x \nabla \x \nn \right)^2       \\
		&+ \frac{1}{2} \left(k_{22} + k_{24}\right) \left[ \Tr \left(\nabla
		\nn\right)^2 - \left(\nabla \cdot \nn \right)^2 \right]\;.
	\end{aligned}
	\label{eq:ffe-oseen}
\end{equation}

\noindent This expression contains the most commonly used elastic terms: splay
($k_{11}$), twist ($k_{22}$) and bend ($k_{33}$). The additional term,
referred to as `saddle-splay', depends on $k_{24}$ and penalizes bidirectional
deformations. {While the prior work demonstrated this {\it can} be measured,
the process requires many more molecules to set up the appropriate anchoring
and boundary conditions for the problem, it is thus significantly slower.
Notably, no measurements of $k_{24}$ exist for $n$CB molecules other than
5CB.} We thus focus here on the standard modes of deformation for which data
can be readily obtained.

The $n$CB molecules are represented using the united atom forcefield
parameterized by Tiberio et al.~\cite{Tiberio2009} After preparing an initial
configuration for each molecule type, the process of obtaining elastic
constant estimates is entirely automated. Each cyanobiphenyl system contains
400 molecules, and is initially prepared by running 100 nanosecond NPT
simulations at 1 atm in Gromacs 5.1.3~\cite{Abraham2015}. The temperatures
examined are between $T_{\mathrm{NI}}$ and $T_{\mathrm{NI}} -
\SI{20}{\kelvin}$. The transition temperature is independently determined by
performing an initial simulation sweep centered on the values reported in
Ref.~\onlinecite{Tiberio2009}, and determining when the equilibrium value of
the nematic order parameter $S$ exceeds 0.1; the measured $T_{\rm NI}$
correspond to those reported in the previous paper. The equilibration time is
considerably less than the 400 ns used for 5CB in our previous work, though a
reasonable length is necessary to allow for elastic constant prediction at
this scale.  A Langevin thermostat and Parrinello--Rahman barostat with
$\tau_{\mathrm{p}} = 5$ picoseconds are used with a time step of 2
femtoseconds for all simulations.

In accordance with Ref.~\onlinecite{Sidky2018}, the average volume at the each
temperature is obtained in the initial round of simulations, and
configurations with this volume are automatically drawn from the completed
$NPT$ trajectories and used to initiate a second round of $NVT$ simulations
with edge restrictions applied using a harmonic restraint having modulus $k =
10^5$ kJ/mol which serves to align the molecules along the $\uvec{z}$ axis.
Four uncorrelated instances are generated at each simulated temperature to
function as independent walkers during the elastic measurement, and enhance
the convergence of the free energy measurement. The simulations at this step
were carried out for \SI{400}{\nano\second}, considerably less than the
\SI{1}{\micro\second} timescale of the previous study~\cite{Sidky2018}. We
arrived at this number through experimentation, but feel that it is entirely
appropriate for most calamitic-type nematogens. Importantly, this step may be
trivially parallelized to increase the speed of convergence to the underlying
free energy landscape and improve the accuracy of the calculations.

The final step involves estimating the elastic constants from a biased
simulation. With edge restrictions in place aligning the molecules along the
$\uvec{z}$ axis, a deformation $\xi$ is applied in the central region according
to the elastic mode desired. We take $\xi$ to be $\partial n_x / \partial x$ for
splay, $\partial n_y / \partial x$ for twist, and $\partial n_z / \partial z$
for bend. Parabolic free energy profiles obtained via basis function
sampling~\cite{Whitmer2014} (BFS) as implemented in SSAGES 0.6~\cite{SSAGES} are
used to extract the final elastic constant as $f = \frac{1}{2} \gamma k_{ii}
\xi^2$ where $\gamma$ is a geometric factor accounting for the finite
restriction and deformation regions. To assist readers in reproducing our
results, we have posted all scripts and Gromacs and SSAGES runfiles in a free
online repository located at \url{https://github.com/hsidky/atomistic_elastics}.

\section*{Results and Discussion}

\begin{figure}[b]
	\begin{center}
	\includegraphics[width=\figurewidth]{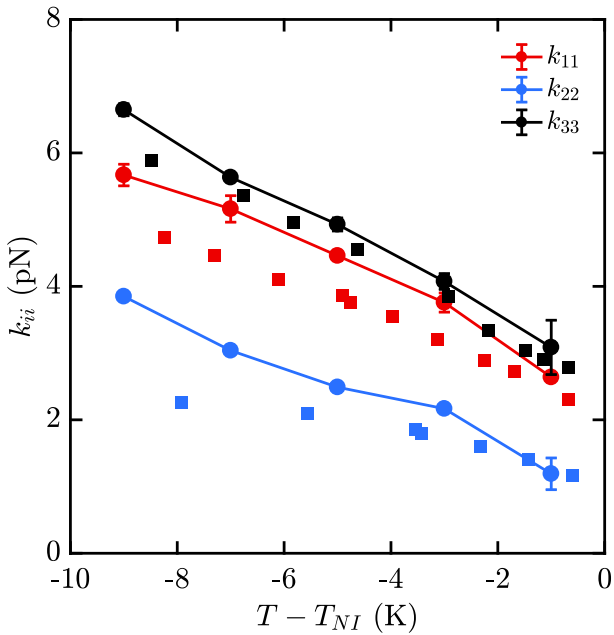}
	\end{center}
	\caption{Predicted bulk elastic constants for 6CB (circles) obtained from
	molecular simulation. Experimental data (squares) obtained from
	Ref.~\onlinecite{Madhusudana1982}. All elastic constants show good general
	agreement, with \bend\ exhibiting the least deviation from experiment and
	\twist\ showing positive deviation, particularly at low temperatures.}
	\label{fig:6cb}
\end{figure}

While data for $n$CB molecules is more sparse than that for 5CB, a few
measurements are available in the literature. We compare all of our
calculations to experimental data from
Refs.~\onlinecite{Madhusudana1982},and~\onlinecite{Chatopadhayay1994}; other
measurements such as those reported in Ref.~\onlinecite{Hakemi1983} were not
used due to substantial deviations from other reported measurements.
Figure~\ref{fig:6cb} shows the elastic constant predictions for 6CB. The
estimated values are in good agreement with experiment across the reported
temperature range. Values for \bend\ and \splay\ fall within the typical range
of variability of experimental measurements reported for 6CB.~\cite{Chen1989a,
Madhusudana1982} Twist elastic constants show a systematic positive drift away
from experimental data as temperature decreases, but remain significantly
smaller than the other moduli, as expected, which results in a negative
deviation of $k_{ii}/k_{22}$. For this measurement, there is minimal noise
observed in our data for $k_{ii}$ with decreasing $T-T_{\rm NI}$, and the
trendline is reasonably smooth and monotonic.

\begin{figure}[t]
	\begin{center}
	\includegraphics[width=\figurewidth]{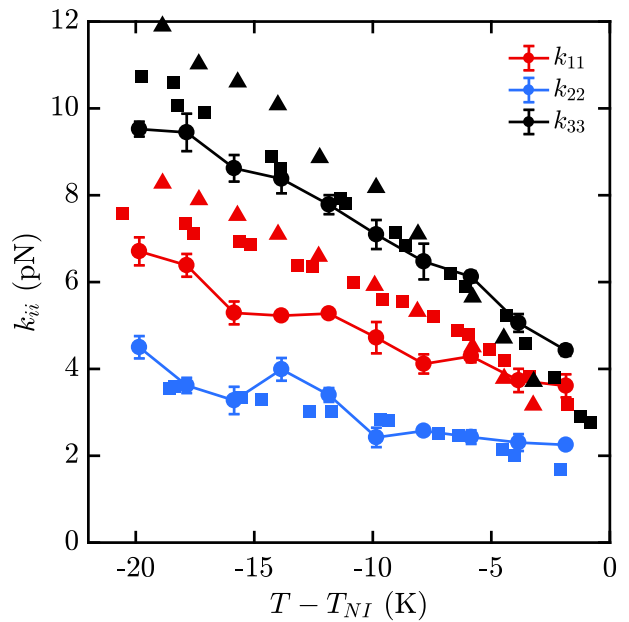}
	\end{center}
	\caption{
	Predicted bulk elastic constants for 7CB (circles) obtained from
	molecular simulation. Experimental data (squares, triangles) obtained from
	Refs.~\onlinecite{Madhusudana1982} and~\onlinecite{Chatopadhayay1994}.
	While elastic constants show excellent agreement with experimental values,
	the estimates exhibit a degree of noise exceeding the bounds indicated by
	the error bars. 
	}
	\label{fig:7cb}
\end{figure}

The elastic constants for 7CB are shown in Figure~\ref{fig:7cb}. They are in
excellent agreement with experiment across a broader temperature range than 6CB,
with only a minor negative deviation of \splay\ at lower temperature. Compared
to 6CB, the \twist\ is exhibits a similar temperature dependence, with
significantly higher \bend\ and \splay\ as temperature decreases. The deviations between experimental and simulation behavior at low temperatures could be due to a form of entropy--enthalpy tradeoff\cite{Chaimovich2010} within the model of Ref.~\onlinecite{Tiberio2009}. These deviations are neverthless small compared to the magnitude of the elastic constants $k_{ii}$. We note that
although quantitative agreement appears to be better than our measurements for 6CB, the measured elastic constants are also noisier,
with data variability exceeding the uncertainty bounds associated with each
individual measurement. This can be understood by clarifying precisely how the
uncertainty estimates were generated, the inherent limitations, and how they an
be addressed. 

As previously mentioned, each free energy estimate and corresponding elastic
constant is obtained using BFS which projects a running estimate of the
partition function onto a truncated basis set. For our purposes, four
independent walkers all contribute to the same free energy estimate to
accelerate convergence. Here the uncertainty is estimated from higher order
terms truncated in the projection, which uses the second order Legendre
polynomial, $P_2$. In other words, the uncertainty is a measure of the
``goodness of fit'' of the discrete biased data to $P_2$. BFS, like many other
free energy sampling methods, is a {\it convergent} algorithm, meaning that
later statistics added into the free energy estimate contribute exponentially
less than early estimates. For slow-relaxing collective variables such as the
elastic deformation used in this work, noisy early estimates can overemphasize
early, noisy data that becomes more difficult to correct over time. It is
possible that this issue can be alleviated utilizing new, neural network-based
methods~\cite{Sidky2018a} which obtain an initial estimate of the surface very
quickly from a small number of counts, or by using ``forgetting'' methods~\cite{Whitmer2014} which
discard a specified fraction of the previous bias history to limit the influence
of early state visits, though these options were not explored in this study.

Ideally, one can also run independent free energy simulations with different
starting configurations and seeds, and to average the resulting statistics into
a single measurement exhibiting upper and lower bounds. At this stage, running a
sufficient number of copies to be confident about the average and error estimate
makes these simulations prohibitively expensive---the ``wall'' time incurred in the simulations as described in the Methods section will increase from $\approx 1$ month
of computing time to $\approx 1$ year or more. An additional potential source
of error is that our elastic constant measurements are carried out in the NVT
ensemble, which introduces additional, albeit less apparently significant,
error, due to our inability to relax the box dimensions in response to applied
orientational stresses. This is a limitation of the deformation collective
variable which does not admit a simple virial expression and thus cannot be run
in NPT simulations. This could be addressed using augmented, joint
density-of-states methods where the box dimensions are additionally sampled via
a coupled Monte-Carlo algorithm.~\cite{Shell2002} Finally, the cumulative
simulation time for systems in this work were a factor of 5 less than those for
5CB~\cite{Sidky2018}, which appears to be a significant factor in overcoming the
initial noisy estimate mentioned. Average wall time was still substantial,
considering the amount of data reported in this work; an cumulative average of
45 days was needed to obtain an elastic constant estimate for each $k_{ii}$ at
each temperature on 24-core Intel Haswell nodes.\footnote{Computers used in this study are 24-core systems comprised of two 12-core Haswell E5-2680v3 processors equipped with a total of 32GB DRAM. Computational time should be significantly less on updated hardware.} Nonetheless, some of the measurement
uncertainty is captured in the reported data, and quantitative accuracy and
trends are not compromised.

\begin{figure}[h]
	\begin{center}
	\includegraphics[width=\figurewidth]{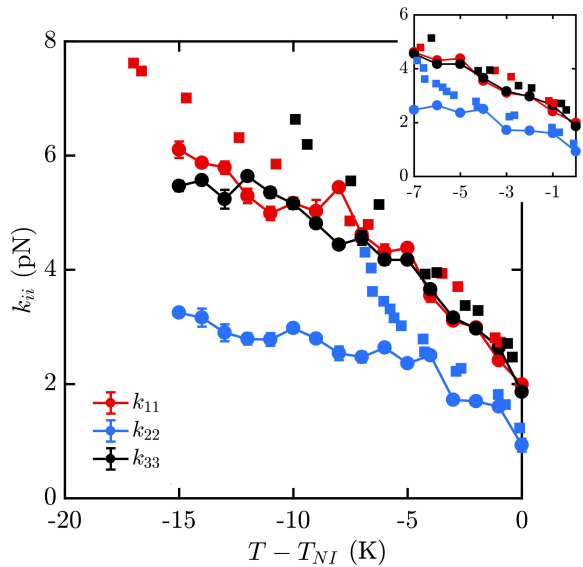}
	\end{center}
	\caption{Predicted bulk elastic constants for 8CB (circles) obtained from
	molecular simulation. Experimental data (squares) obtained from
	Ref.~\onlinecite{Madhusudana1982}. The simulated elastic constants are in
	excellent agreement with experimental data within the nematic range (see
	inset) and show $k_{11} \approx k_{33}$. However, at lower temperatures we
	do not observe a divergence in \twist and \bend\ commensurate with the
	formation of a smectic phase. This is due to the suppression and disruption
	of layered ordering caused by the NVT free energy sampling process and 
	strong finite size effects of smectic ordering.}
	\label{fig:8cb}
\end{figure}

Figure~\ref{fig:8cb} shows the elastic moduli for 8CB. It is well known that
8CB exhibits a smectic-A--nematic transition at approximately \SI{7}{\kelvin}
below $T_{NI}$. which results in a divergence of \twist\ and
\bend.~\cite{Madhusudana1982} The formation of layers along the nematic axis
permits only deformations which are curl-free.~\cite{DeGennes1972} This gives
rise to a power-law scaling relation with a critical exponent of $\nu \approx
0.67$.~\cite{Hakemi1989} Additionally, the model used in this study has been
shown to accurately capture the both transition temperatures and smectic
layering behavior~\cite{Palermo2013}, although with strong finite size
dependence. Our results across the nematic phase, shown in the
Figure~\ref{fig:8cb} inset, are an excellent match to the experimental
results. However, as the experimental \twist\ begins to diverge through the
smectic-nematic transition, we see that our simulated results do not exhibit
the same behavior. Plotted across the entire range of experimental
measurements, it is clear that our measurements are not indicative of the
presence of a smectic phase within our small samples.

This apparent discrepancy is due to a number of effects. The first is a strong
finite size dependence which has already been reported
previously.~\cite{Palermo2013} Compared to larger systems, those containing
fewer than $N = 1000$ mesogens, which is our case, exhibit a considerable
discrepancy in smectic ordering.~\cite{Palermo2013} Similar behaviors were
previously observed in combined phase and elastic property simulations of Gay
--Berne models.\cite{Joshi2014} The second is that we force the alignment of
the 400 mesogens along the $\uvec{z}$ axis which may not be the preferred
orientation of layers in a fixed-size periodic box. Thirdly, deformations to
the mesogens are applied without considering the presence of smectic layering.
Finally, simulations are carried out in the NVT ensemble which does not allow
collective orientational relaxation. All of these factors contribute to the
suppression and disruption of smectic layering which would manifest in the
divergent elasticities absent from our predictions. To properly accommodate
and characterize natural smectic layering a semi-isotropic barostat would be
necessary, allowing the box dimension ratios to adopt those corresponding to
layer separation. The difficulties associated with this within the context of
our elastic measurements have been discussed above. In principle, however, the
methods we present may be extended to study this intriguing behavior in
systems on the order of $N=3000$ mesogens~\cite{Palermo2013}.

\section*{Conclusion}

In this work we have estimated the elastic constants of the cyanobiphenyls
6CB, 7CB, and 8CB from biased molecular simulation using an entirely automated
procedure.    We show that the simulated elastic moduli are generally in good
agreement with reported  experimental values. Difficulties arise in generating
meaningful uncertainties and dealing  with the onset of the smectic phase for
8CB. Both improving the prediction accuracy and  applying this methodology on
a large scale are only possible if the cost of elasticity  measurements is
substantially reduced. Currently, up to 4 $\mu s$ of total simulation runtime
is required to generate a estimate of a single elastic mode at a single
temperature.  

Developing newer or better optimized sampling algorithms may
significantly reduce this time  and open the door to {\it in silico} liquid
crystal engineering. This would necessitate some means of attenuating early-time 
noise present in elastic measurements, and perhaps a simplification of the 
order-parameter calculation which is computationally expensive and involves 
third-order tensor algebra. As presented, the methodology developed
is capable of producing accurate elastic constants and merits further study 
to elucidate the precise limit of efficiency while retaining measurement 
fidelity. Further, a crucial next step involves moving beyond cyanobiphenyls and examine atomistic models of the myriad liquid crystal phases, from molecules of historical interest, such as PAA and MBBA\cite{Haller1972, deJeu1976} to more exotic bent-core molecules and models exhibiting chiral\cite{Dierking2014} or twist-bend phases.\cite{Chen2013} Improving the breadth, speed, and accuracy of these techniques will enable their use in computational screening of new mesogenic compounds.

\section*{Code Availability}
All scripts and information necessary to run the examples contained in this article are posted at \url{https://github.com/hsidky/atomistic_elastics}.

\section*{Acknowledgement}
HS acknowledges support from the National Science Foundation Graduate Research
Fellowship Program (NSF-GRFP). JKW acknowledges the support of MICCoM, the
Midwest Center for Computational Materials, as part of the Computational
Materials Sciences Program funded by the U.S.  Department of Energy, Office of
Science, Basic Energy Sciences, Materials Sciences  and Engineering Division,
for the development of algorithms and codes used within this work. HS and JKW
additionally acknowledge computational resources at the Notre Dame Center for
Research Computing (CRC).

\bibliography{references}

\begin{thebibliography}{39}
\expandafter\ifx\csname natexlab\endcsname\relax\def\natexlab#1{#1}\fi
\expandafter\ifx\csname bibnamefont\endcsname\relax
  \def\bibnamefont#1{#1}\fi
\expandafter\ifx\csname bibfnamefont\endcsname\relax
  \def\bibfnamefont#1{#1}\fi
\expandafter\ifx\csname citenamefont\endcsname\relax
  \def\citenamefont#1{#1}\fi
\expandafter\ifx\csname url\endcsname\relax
  \def\url#1{\texttt{#1}}\fi
\expandafter\ifx\csname urlprefix\endcsname\relax\def\urlprefix{URL }\fi
\providecommand{\bibinfo}[2]{#2}
\providecommand{\eprint}[2][]{\url{#2}}

\bibitem[{\citenamefont{Kleman and Lavrentovich}(2003)}]{SoftMatterPhys}
\bibinfo{author}{\bibfnamefont{M.}~\bibnamefont{Kleman}} \bibnamefont{and}
  \bibinfo{author}{\bibfnamefont{O.~D.} \bibnamefont{Lavrentovich}},
  \emph{\bibinfo{title}{Soft Matter Physics: An Introduction}}
  (\bibinfo{publisher}{Springer-Verlag New York}, \bibinfo{year}{2003}),
  \bibinfo{edition}{1st} ed., ISBN \bibinfo{isbn}{978-0-387-95267-3}.

\bibitem[{\citenamefont{Gray and Kelly}(1999)}]{Gray1999}
\bibinfo{author}{\bibfnamefont{G.~W.} \bibnamefont{Gray}} \bibnamefont{and}
  \bibinfo{author}{\bibfnamefont{S.~M.} \bibnamefont{Kelly}},
  \bibinfo{journal}{Journal of Materials Chemistry}
  \textbf{\bibinfo{volume}{9}}, \bibinfo{pages}{2037} (\bibinfo{year}{1999}).

\bibitem[{\citenamefont{Szilv{\'{a}}si
  et~al.}(2017)\citenamefont{Szilv{\'{a}}si, Roling, Yu, Rai, Choi, Twieg,
  Mavrikakis, and Abbott}}]{tibor2017}
\bibinfo{author}{\bibfnamefont{T.}~\bibnamefont{Szilv{\'{a}}si}},
  \bibinfo{author}{\bibfnamefont{L.~T.} \bibnamefont{Roling}},
  \bibinfo{author}{\bibfnamefont{H.}~\bibnamefont{Yu}},
  \bibinfo{author}{\bibfnamefont{P.}~\bibnamefont{Rai}},
  \bibinfo{author}{\bibfnamefont{S.}~\bibnamefont{Choi}},
  \bibinfo{author}{\bibfnamefont{R.~J.} \bibnamefont{Twieg}},
  \bibinfo{author}{\bibfnamefont{M.}~\bibnamefont{Mavrikakis}},
  \bibnamefont{and} \bibinfo{author}{\bibfnamefont{N.~L.}
  \bibnamefont{Abbott}}, \bibinfo{journal}{Chemistry of Materials}
  \textbf{\bibinfo{volume}{29}}, \bibinfo{pages}{3563} (\bibinfo{year}{2017}),
  \urlprefix\url{https://doi.org/10.1021/acs.chemmater.6b05430}.

\bibitem[{\citenamefont{Babakhanova et~al.}(2018)\citenamefont{Babakhanova,
  Turiv, Guo, Hendrikx, Wei, Schenning, Broer, and
  Lavrentovich}}]{Babakhanova2018}
\bibinfo{author}{\bibfnamefont{G.}~\bibnamefont{Babakhanova}},
  \bibinfo{author}{\bibfnamefont{T.}~\bibnamefont{Turiv}},
  \bibinfo{author}{\bibfnamefont{Y.}~\bibnamefont{Guo}},
  \bibinfo{author}{\bibfnamefont{M.}~\bibnamefont{Hendrikx}},
  \bibinfo{author}{\bibfnamefont{Q.-H.} \bibnamefont{Wei}},
  \bibinfo{author}{\bibfnamefont{A.~P. H.~J.} \bibnamefont{Schenning}},
  \bibinfo{author}{\bibfnamefont{D.~J.} \bibnamefont{Broer}}, \bibnamefont{and}
  \bibinfo{author}{\bibfnamefont{O.~D.} \bibnamefont{Lavrentovich}},
  \bibinfo{journal}{Nature Communications} \textbf{\bibinfo{volume}{9}},
  \bibinfo{pages}{456} (\bibinfo{year}{2018}), ISSN \bibinfo{issn}{2041-1723},
  \urlprefix\url{https://doi.org/10.1038/s41467-018-02895-9}.

\bibitem[{\citenamefont{Wang et~al.}(2015)\citenamefont{Wang, Miller,
  Bukusoglu, de~Pablo, and Abbott}}]{Wang2015}
\bibinfo{author}{\bibfnamefont{X.}~\bibnamefont{Wang}},
  \bibinfo{author}{\bibfnamefont{D.~S.} \bibnamefont{Miller}},
  \bibinfo{author}{\bibfnamefont{E.}~\bibnamefont{Bukusoglu}},
  \bibinfo{author}{\bibfnamefont{J.~J.} \bibnamefont{de~Pablo}},
  \bibnamefont{and} \bibinfo{author}{\bibfnamefont{N.~L.}
  \bibnamefont{Abbott}}, \bibinfo{journal}{Nature Materials}
  \textbf{\bibinfo{volume}{15}}, \bibinfo{pages}{106} (\bibinfo{year}{2015}),
  ISSN \bibinfo{issn}{1476-1122}.

\bibitem[{\citenamefont{Palermo et~al.}(2013)\citenamefont{Palermo, Pizzirusso,
  Muccioli, and Zannoni}}]{Palermo2013}
\bibinfo{author}{\bibfnamefont{M.~F.} \bibnamefont{Palermo}},
  \bibinfo{author}{\bibfnamefont{A.}~\bibnamefont{Pizzirusso}},
  \bibinfo{author}{\bibfnamefont{L.}~\bibnamefont{Muccioli}}, \bibnamefont{and}
  \bibinfo{author}{\bibfnamefont{C.}~\bibnamefont{Zannoni}},
  \bibinfo{journal}{The Journal of Chemical Physics}
  \textbf{\bibinfo{volume}{138}}, \bibinfo{pages}{204901}
  (\bibinfo{year}{2013}), ISSN \bibinfo{issn}{0021-9606}.

\bibitem[{\citenamefont{Ramezani-Dakhel
  et~al.}(2017)\citenamefont{Ramezani-Dakhel, Sadati, Rahimi,
  Ramirez-Hernandez, Roux, and de~Pablo}}]{Ramezani-Dakhel2017}
\bibinfo{author}{\bibfnamefont{H.}~\bibnamefont{Ramezani-Dakhel}},
  \bibinfo{author}{\bibfnamefont{M.}~\bibnamefont{Sadati}},
  \bibinfo{author}{\bibfnamefont{M.}~\bibnamefont{Rahimi}},
  \bibinfo{author}{\bibfnamefont{A.}~\bibnamefont{Ramirez-Hernandez}},
  \bibinfo{author}{\bibfnamefont{B.}~\bibnamefont{Roux}}, \bibnamefont{and}
  \bibinfo{author}{\bibfnamefont{J.~J.} \bibnamefont{de~Pablo}},
  \bibinfo{journal}{Journal of Chemical Theory and Computation}
  \textbf{\bibinfo{volume}{13}}, \bibinfo{pages}{237} (\bibinfo{year}{2017}),
  ISSN \bibinfo{issn}{1549-9618}.

\bibitem[{\citenamefont{Sidky et~al.}(2018{\natexlab{a}})\citenamefont{Sidky,
  de~Pablo, and Whitmer}}]{Sidky2018}
\bibinfo{author}{\bibfnamefont{H.}~\bibnamefont{Sidky}},
  \bibinfo{author}{\bibfnamefont{J.~J.} \bibnamefont{de~Pablo}},
  \bibnamefont{and} \bibinfo{author}{\bibfnamefont{J.~K.}
  \bibnamefont{Whitmer}}, \bibinfo{journal}{Physical Review Letters}
  \textbf{\bibinfo{volume}{120}}, \bibinfo{pages}{107801}
  (\bibinfo{year}{2018}{\natexlab{a}}).

\bibitem[{\citenamefont{Allender et~al.}(1991)\citenamefont{Allender, Crawford,
  and Doane}}]{Allender1991}
\bibinfo{author}{\bibfnamefont{D.~W.} \bibnamefont{Allender}},
  \bibinfo{author}{\bibfnamefont{G.~P.} \bibnamefont{Crawford}},
  \bibnamefont{and} \bibinfo{author}{\bibfnamefont{J.~W.} \bibnamefont{Doane}},
  \bibinfo{journal}{Physical Review Letters} \textbf{\bibinfo{volume}{67}},
  \bibinfo{pages}{1442} (\bibinfo{year}{1991}), ISSN \bibinfo{issn}{0031-9007},
  \urlprefix\url{http://link.aps.org/doi/10.1103/PhysRevLett.67.1442}.

\bibitem[{\citenamefont{Sparavigna et~al.}(1994)\citenamefont{Sparavigna,
  Lavrentovich, and Strigazzi}}]{Sparavigna1994}
\bibinfo{author}{\bibfnamefont{A.}~\bibnamefont{Sparavigna}},
  \bibinfo{author}{\bibfnamefont{O.~D.} \bibnamefont{Lavrentovich}},
  \bibnamefont{and}
  \bibinfo{author}{\bibfnamefont{A.}~\bibnamefont{Strigazzi}},
  \bibinfo{journal}{Physical Review E} \textbf{\bibinfo{volume}{49}},
  \bibinfo{pages}{1344} (\bibinfo{year}{1994}), ISSN \bibinfo{issn}{1063651X}.

\bibitem[{\citenamefont{Polak et~al.}(1994)\citenamefont{Polak, Crawford,
  Kostival, Doane, and \ifmmode~\check{Z}\else \v{Z}\fi{}umer}}]{Polak1994}
\bibinfo{author}{\bibfnamefont{R.~D.} \bibnamefont{Polak}},
  \bibinfo{author}{\bibfnamefont{G.~P.} \bibnamefont{Crawford}},
  \bibinfo{author}{\bibfnamefont{B.~C.} \bibnamefont{Kostival}},
  \bibinfo{author}{\bibfnamefont{J.~W.} \bibnamefont{Doane}}, \bibnamefont{and}
  \bibinfo{author}{\bibfnamefont{S.}~\bibnamefont{\ifmmode~\check{Z}\else
  \v{Z}\fi{}umer}}, \bibinfo{journal}{Physical Review E}
  \textbf{\bibinfo{volume}{49}}, \bibinfo{pages}{R978} (\bibinfo{year}{1994}),
  \urlprefix\url{https://link.aps.org/doi/10.1103/PhysRevE.49.R978}.

\bibitem[{\citenamefont{Pairam et~al.}(2013)\citenamefont{Pairam, Vallamkondu,
  Koning, van Zuiden, Ellis, Bates, Vitelli, and
  Fernandez-Nieves}}]{Pairam2013}
\bibinfo{author}{\bibfnamefont{E.}~\bibnamefont{Pairam}},
  \bibinfo{author}{\bibfnamefont{J.}~\bibnamefont{Vallamkondu}},
  \bibinfo{author}{\bibfnamefont{V.}~\bibnamefont{Koning}},
  \bibinfo{author}{\bibfnamefont{B.~C.} \bibnamefont{van Zuiden}},
  \bibinfo{author}{\bibfnamefont{P.~W.} \bibnamefont{Ellis}},
  \bibinfo{author}{\bibfnamefont{M.~A.} \bibnamefont{Bates}},
  \bibinfo{author}{\bibfnamefont{V.}~\bibnamefont{Vitelli}}, \bibnamefont{and}
  \bibinfo{author}{\bibfnamefont{A.}~\bibnamefont{Fernandez-Nieves}},
  \bibinfo{journal}{Proceedings of the National Academy of Sciences of the
  United States of America} \textbf{\bibinfo{volume}{110}},
  \bibinfo{pages}{9295} (\bibinfo{year}{2013}), ISSN \bibinfo{issn}{1091-6490},
  \urlprefix\url{http://www.pnas.org/content/110/23/9295}.

\bibitem[{\citenamefont{Davidson et~al.}(2015)\citenamefont{Davidson, Kang,
  Jeong, Still, Collings, Lubensky, and Yodh}}]{Davidson2015}
\bibinfo{author}{\bibfnamefont{Z.~S.} \bibnamefont{Davidson}},
  \bibinfo{author}{\bibfnamefont{L.}~\bibnamefont{Kang}},
  \bibinfo{author}{\bibfnamefont{J.}~\bibnamefont{Jeong}},
  \bibinfo{author}{\bibfnamefont{T.}~\bibnamefont{Still}},
  \bibinfo{author}{\bibfnamefont{P.~J.} \bibnamefont{Collings}},
  \bibinfo{author}{\bibfnamefont{T.~C.} \bibnamefont{Lubensky}},
  \bibnamefont{and} \bibinfo{author}{\bibfnamefont{A.~G.} \bibnamefont{Yodh}},
  \bibinfo{journal}{Physical Review E} \textbf{\bibinfo{volume}{91}},
  \bibinfo{pages}{050501} (\bibinfo{year}{2015}), ISSN
  \bibinfo{issn}{15502376},
  \urlprefix\url{http://link.aps.org/doi/10.1103/PhysRevE.91.050501}.

\bibitem[{\citenamefont{Wang et~al.}(2016)\citenamefont{Wang, Miller,
  Bukusoglu, de~Pablo, and Abbott}}]{Wang2016b}
\bibinfo{author}{\bibfnamefont{X.}~\bibnamefont{Wang}},
  \bibinfo{author}{\bibfnamefont{D.~S.} \bibnamefont{Miller}},
  \bibinfo{author}{\bibfnamefont{E.}~\bibnamefont{Bukusoglu}},
  \bibinfo{author}{\bibfnamefont{J.~J.} \bibnamefont{de~Pablo}},
  \bibnamefont{and} \bibinfo{author}{\bibfnamefont{N.~L.}
  \bibnamefont{Abbott}}, \bibinfo{journal}{Nature Materials}
  \textbf{\bibinfo{volume}{15}}, \bibinfo{pages}{106} (\bibinfo{year}{2016}).

\bibitem[{\citenamefont{Eimura et~al.}(2016)\citenamefont{Eimura, Miller, Wang,
  Abbott, and Kato}}]{Eimura2016}
\bibinfo{author}{\bibfnamefont{H.}~\bibnamefont{Eimura}},
  \bibinfo{author}{\bibfnamefont{D.~S.} \bibnamefont{Miller}},
  \bibinfo{author}{\bibfnamefont{X.}~\bibnamefont{Wang}},
  \bibinfo{author}{\bibfnamefont{N.~L.} \bibnamefont{Abbott}},
  \bibnamefont{and} \bibinfo{author}{\bibfnamefont{T.}~\bibnamefont{Kato}},
  \bibinfo{journal}{Chemistry of Materials} \textbf{\bibinfo{volume}{28}},
  \bibinfo{pages}{1170} (\bibinfo{year}{2016}),
  \eprint{http://dx.doi.org/10.1021/acs.chemmater.5b04736}.

\bibitem[{\citenamefont{Madhusudana and Pratibha}(1982)}]{Madhusudana1982}
\bibinfo{author}{\bibfnamefont{N.~V.} \bibnamefont{Madhusudana}}
  \bibnamefont{and} \bibinfo{author}{\bibfnamefont{R.}~\bibnamefont{Pratibha}},
  \bibinfo{journal}{Molecular Crystals and Liquid Crystals}
  \textbf{\bibinfo{volume}{89}}, \bibinfo{pages}{249} (\bibinfo{year}{1982}),
  ISSN \bibinfo{issn}{0026-8941},
  \urlprefix\url{http://www.tandfonline.com/doi/abs/10.1080/00268948208074481}.

\bibitem[{\citenamefont{Hakemi et~al.}(1983)\citenamefont{Hakemi, Jagodzinski,
  and DuPre}}]{Hakemi1983}
\bibinfo{author}{\bibfnamefont{H.}~\bibnamefont{Hakemi}},
  \bibinfo{author}{\bibfnamefont{E.~F.} \bibnamefont{Jagodzinski}},
  \bibnamefont{and} \bibinfo{author}{\bibfnamefont{D.~B.} \bibnamefont{DuPre}},
  \bibinfo{journal}{The Journal of Chemical Physics}
  \textbf{\bibinfo{volume}{78}}, \bibinfo{pages}{1513} (\bibinfo{year}{1983}),
  ISSN \bibinfo{issn}{00219606}.

\bibitem[{\citenamefont{Chen et~al.}(1989)\citenamefont{Chen, Takezoe, and
  Fukuda}}]{Chen1989a}
\bibinfo{author}{\bibfnamefont{G.-P.} \bibnamefont{Chen}},
  \bibinfo{author}{\bibfnamefont{H.}~\bibnamefont{Takezoe}}, \bibnamefont{and}
  \bibinfo{author}{\bibfnamefont{A.}~\bibnamefont{Fukuda}},
  \bibinfo{journal}{Liquid Crystals} \textbf{\bibinfo{volume}{5}},
  \bibinfo{pages}{341} (\bibinfo{year}{1989}), ISSN \bibinfo{issn}{0267-8292}.

\bibitem[{\citenamefont{Chatopadhayay and Roy}(1994)}]{Chatopadhayay1994}
\bibinfo{author}{\bibfnamefont{P.}~\bibnamefont{Chatopadhayay}}
  \bibnamefont{and} \bibinfo{author}{\bibfnamefont{S.~K.} \bibnamefont{Roy}},
  \bibinfo{journal}{Molecular Crystals and Liquid Crystals Science and
  Technology. Section A. Molecular Crystals and Liquid Crystals}
  \textbf{\bibinfo{volume}{257}}, \bibinfo{pages}{89} (\bibinfo{year}{1994}),
  ISSN \bibinfo{issn}{1058-725X}.

\bibitem[{\citenamefont{Tiberio et~al.}(2009)\citenamefont{Tiberio, Muccioli,
  Berardi, and Zannoni}}]{Tiberio2009}
\bibinfo{author}{\bibfnamefont{G.}~\bibnamefont{Tiberio}},
  \bibinfo{author}{\bibfnamefont{L.}~\bibnamefont{Muccioli}},
  \bibinfo{author}{\bibfnamefont{R.}~\bibnamefont{Berardi}}, \bibnamefont{and}
  \bibinfo{author}{\bibfnamefont{C.}~\bibnamefont{Zannoni}},
  \bibinfo{journal}{ChemPhysChem} \textbf{\bibinfo{volume}{10}},
  \bibinfo{pages}{125} (\bibinfo{year}{2009}).

\bibitem[{\citenamefont{Boyd and Wilson}(2015)}]{Boyd2015}
\bibinfo{author}{\bibfnamefont{N.~J.} \bibnamefont{Boyd}} \bibnamefont{and}
  \bibinfo{author}{\bibfnamefont{M.~R.} \bibnamefont{Wilson}},
  \bibinfo{journal}{Physical Chemistry Chemical Physics}
  \textbf{\bibinfo{volume}{17}}, \bibinfo{pages}{24851} (\bibinfo{year}{2015}),
  ISSN \bibinfo{issn}{1463-9076}.

\bibitem[{\citenamefont{Cacelli et~al.}(2007)\citenamefont{Cacelli, {De
  Gaetani}, Prampolini, and Tani}}]{Cacelli2007}
\bibinfo{author}{\bibfnamefont{I.}~\bibnamefont{Cacelli}},
  \bibinfo{author}{\bibfnamefont{L.}~\bibnamefont{{De Gaetani}}},
  \bibinfo{author}{\bibfnamefont{G.}~\bibnamefont{Prampolini}},
  \bibnamefont{and} \bibinfo{author}{\bibfnamefont{A.}~\bibnamefont{Tani}},
  \bibinfo{journal}{The Journal of Physical Chemistry B}
  \textbf{\bibinfo{volume}{111}}, \bibinfo{pages}{2130} (\bibinfo{year}{2007}),
  ISSN \bibinfo{issn}{1520-6106},
  \urlprefix\url{http://pubs.acs.org/doi/abs/10.1021/jp065806l}.

\bibitem[{\citenamefont{Cacelli et~al.}(2009)\citenamefont{Cacelli, Lami, and
  Prampolini}}]{Cacelli2009}
\bibinfo{author}{\bibfnamefont{I.}~\bibnamefont{Cacelli}},
  \bibinfo{author}{\bibfnamefont{C.~F.} \bibnamefont{Lami}}, \bibnamefont{and}
  \bibinfo{author}{\bibfnamefont{G.}~\bibnamefont{Prampolini}},
  \bibinfo{journal}{Journal of Computational Chemistry}
  \textbf{\bibinfo{volume}{30}}, \bibinfo{pages}{366} (\bibinfo{year}{2009}),
  ISSN \bibinfo{issn}{01928651}.

\bibitem[{\citenamefont{Gray and Harrison}(1971)}]{Gray1971}
\bibinfo{author}{\bibfnamefont{G.~W.} \bibnamefont{Gray}} \bibnamefont{and}
  \bibinfo{author}{\bibfnamefont{K.~J.} \bibnamefont{Harrison}},
  \bibinfo{journal}{Symp. Faraday Soc.} \textbf{\bibinfo{volume}{5}},
  \bibinfo{pages}{54} (\bibinfo{year}{1971}),
  \urlprefix\url{http://dx.doi.org/10.1039/SF9710500054}.

\bibitem[{\citenamefont{Stewart}(2004)}]{Stewart2004}
\bibinfo{author}{\bibfnamefont{I.~W.} \bibnamefont{Stewart}},
  \emph{\bibinfo{title}{The static and dynamic continuum theory of liquid
  crystals: a mathematical introduction}} (\bibinfo{publisher}{Taylor \&
  Francis}, \bibinfo{address}{London}, \bibinfo{year}{2004}).

\bibitem[{\citenamefont{Abraham et~al.}(2015)\citenamefont{Abraham, Murtola,
  Schulz, P{\'{a}}ll, Smith, Hess, and Lindahl}}]{Abraham2015}
\bibinfo{author}{\bibfnamefont{M.~J.} \bibnamefont{Abraham}},
  \bibinfo{author}{\bibfnamefont{T.}~\bibnamefont{Murtola}},
  \bibinfo{author}{\bibfnamefont{R.}~\bibnamefont{Schulz}},
  \bibinfo{author}{\bibfnamefont{S.}~\bibnamefont{P{\'{a}}ll}},
  \bibinfo{author}{\bibfnamefont{J.~C.} \bibnamefont{Smith}},
  \bibinfo{author}{\bibfnamefont{B.}~\bibnamefont{Hess}}, \bibnamefont{and}
  \bibinfo{author}{\bibfnamefont{E.}~\bibnamefont{Lindahl}},
  \bibinfo{journal}{SoftwareX} \textbf{\bibinfo{volume}{1–2}},
  \bibinfo{pages}{19} (\bibinfo{year}{2015}), ISSN \bibinfo{issn}{2352-7110},
  \urlprefix\url{http://www.sciencedirect.com/science/article/pii/S2352711015000059}.

\bibitem[{\citenamefont{Whitmer et~al.}(2014)\citenamefont{Whitmer, Chiu,
  Joshi, and de~Pablo}}]{Whitmer2014}
\bibinfo{author}{\bibfnamefont{J.~K.} \bibnamefont{Whitmer}},
  \bibinfo{author}{\bibfnamefont{C.-c.} \bibnamefont{Chiu}},
  \bibinfo{author}{\bibfnamefont{A.~A.} \bibnamefont{Joshi}}, \bibnamefont{and}
  \bibinfo{author}{\bibfnamefont{J.~J.} \bibnamefont{de~Pablo}},
  \bibinfo{journal}{Physical Review Letters} \textbf{\bibinfo{volume}{113}},
  \bibinfo{pages}{190602} (\bibinfo{year}{2014}), ISSN
  \bibinfo{issn}{0031-9007},
  \urlprefix\url{https://link.aps.org/doi/10.1103/PhysRevLett.113.190602}.

\bibitem[{\citenamefont{Sidky et~al.}(2018{\natexlab{b}})\citenamefont{Sidky,
  Col{\'{o}}n, Helfferich, Sikora, Bezik, Chu, Giberti, Guo, Jiang, Lequieu
  et~al.}}]{SSAGES}
\bibinfo{author}{\bibfnamefont{H.}~\bibnamefont{Sidky}},
  \bibinfo{author}{\bibfnamefont{Y.~J.} \bibnamefont{Col{\'{o}}n}},
  \bibinfo{author}{\bibfnamefont{J.}~\bibnamefont{Helfferich}},
  \bibinfo{author}{\bibfnamefont{B.~J.} \bibnamefont{Sikora}},
  \bibinfo{author}{\bibfnamefont{C.}~\bibnamefont{Bezik}},
  \bibinfo{author}{\bibfnamefont{W.}~\bibnamefont{Chu}},
  \bibinfo{author}{\bibfnamefont{F.}~\bibnamefont{Giberti}},
  \bibinfo{author}{\bibfnamefont{A.~Z.} \bibnamefont{Guo}},
  \bibinfo{author}{\bibfnamefont{X.}~\bibnamefont{Jiang}},
  \bibinfo{author}{\bibfnamefont{J.}~\bibnamefont{Lequieu}},
  \bibnamefont{et~al.}, \bibinfo{journal}{The Journal of Chemical Physics}
  \textbf{\bibinfo{volume}{148}}, \bibinfo{pages}{044104}
  (\bibinfo{year}{2018}{\natexlab{b}}), ISSN \bibinfo{issn}{0021-9606},
  \urlprefix\url{http://aip.scitation.org/doi/10.1063/1.5008853}.

\bibitem[{\citenamefont{Chaimovich and Shell}(2010)}]{Chaimovich2010}
\bibinfo{author}{\bibfnamefont{A.}~\bibnamefont{Chaimovich}} \bibnamefont{and}
  \bibinfo{author}{\bibfnamefont{M.~S.} \bibnamefont{Shell}},
  \bibinfo{journal}{Phys. Rev. E} \textbf{\bibinfo{volume}{81}},
  \bibinfo{pages}{060104} (\bibinfo{year}{2010}),
  \urlprefix\url{https://link.aps.org/doi/10.1103/PhysRevE.81.060104}.

\bibitem[{\citenamefont{Sidky and Whitmer}(2018)}]{Sidky2018a}
\bibinfo{author}{\bibfnamefont{H.}~\bibnamefont{Sidky}} \bibnamefont{and}
  \bibinfo{author}{\bibfnamefont{J.~K.} \bibnamefont{Whitmer}},
  \bibinfo{journal}{The Journal of Chemical Physics}
  \textbf{\bibinfo{volume}{148}}, \bibinfo{pages}{104111}
  (\bibinfo{year}{2018}), ISSN \bibinfo{issn}{0021-9606}.

\bibitem[{\citenamefont{Shell et~al.}(2002)\citenamefont{Shell, Debenedetti,
  and Panagiotopoulos}}]{Shell2002}
\bibinfo{author}{\bibfnamefont{M.~S.} \bibnamefont{Shell}},
  \bibinfo{author}{\bibfnamefont{P.~G.} \bibnamefont{Debenedetti}},
  \bibnamefont{and} \bibinfo{author}{\bibfnamefont{A.~Z.}
  \bibnamefont{Panagiotopoulos}}, \bibinfo{journal}{Physical Review E}
  \textbf{\bibinfo{volume}{66}}, \bibinfo{pages}{056703}
  (\bibinfo{year}{2002}), ISSN \bibinfo{issn}{1063651X}.

\bibitem[{Note1()}]{Note1}
Note1, \bibinfo{note}{computers used in this study are 24-core systems
  comprised of two 12-core Haswell E5-2680v3 processors equipped with a total
  of 32GB DRAM. Computational time should be significantly less on updated
  hardware.}

\bibitem[{\citenamefont{de~Gennes}(1972)}]{DeGennes1972}
\bibinfo{author}{\bibfnamefont{P.}~\bibnamefont{de~Gennes}},
  \bibinfo{journal}{Solid State Communications} \textbf{\bibinfo{volume}{10}},
  \bibinfo{pages}{753} (\bibinfo{year}{1972}), ISSN \bibinfo{issn}{00381098}.

\bibitem[{\citenamefont{Hakemi}(1989)}]{Hakemi1989}
\bibinfo{author}{\bibfnamefont{H.}~\bibnamefont{Hakemi}},
  \bibinfo{journal}{Liquid Crystals} \textbf{\bibinfo{volume}{5}},
  \bibinfo{pages}{327} (\bibinfo{year}{1989}), ISSN \bibinfo{issn}{0267-8292}.

\bibitem[{\citenamefont{Joshi et~al.}(2014)\citenamefont{Joshi, Whitmer,
  Guzm{\'{a}}n, Abbott, and de~Pablo}}]{Joshi2014}
\bibinfo{author}{\bibfnamefont{A.~A.} \bibnamefont{Joshi}},
  \bibinfo{author}{\bibfnamefont{J.~K.} \bibnamefont{Whitmer}},
  \bibinfo{author}{\bibfnamefont{O.}~\bibnamefont{Guzm{\'{a}}n}},
  \bibinfo{author}{\bibfnamefont{N.~L.} \bibnamefont{Abbott}},
  \bibnamefont{and} \bibinfo{author}{\bibfnamefont{J.~J.}
  \bibnamefont{de~Pablo}}, \bibinfo{journal}{Soft matter}
  \textbf{\bibinfo{volume}{10}}, \bibinfo{pages}{882} (\bibinfo{year}{2014}),
  ISSN \bibinfo{issn}{1744-6848},
  \urlprefix\url{http://www.ncbi.nlm.nih.gov/pubmed/24837037}.

\bibitem[{\citenamefont{Haller}(1972)}]{Haller1972}
\bibinfo{author}{\bibfnamefont{I.}~\bibnamefont{Haller}}, \bibinfo{journal}{The
  Journal of Chemical Physics} \textbf{\bibinfo{volume}{57}},
  \bibinfo{pages}{1400} (\bibinfo{year}{1972}),
  \eprint{https://doi.org/10.1063/1.1678416},
  \urlprefix\url{https://doi.org/10.1063/1.1678416}.

\bibitem[{\citenamefont{Jeu et~al.}(1976)\citenamefont{Jeu, Claassen, and
  Spruijt}}]{deJeu1976}
\bibinfo{author}{\bibfnamefont{W.~H.~D.} \bibnamefont{Jeu}},
  \bibinfo{author}{\bibfnamefont{W.~A.~P.} \bibnamefont{Claassen}},
  \bibnamefont{and} \bibinfo{author}{\bibfnamefont{A.~M.~J.}
  \bibnamefont{Spruijt}}, \bibinfo{journal}{Molecular Crystals and Liquid
  Crystals} \textbf{\bibinfo{volume}{37}}, \bibinfo{pages}{269}
  (\bibinfo{year}{1976}), \eprint{https://doi.org/10.1080/15421407608084362},
  \urlprefix\url{https://doi.org/10.1080/15421407608084362}.

\bibitem[{\citenamefont{Dierking}(2014)}]{Dierking2014}
\bibinfo{author}{\bibfnamefont{I.}~\bibnamefont{Dierking}},
  \bibinfo{journal}{Symmetry} \textbf{\bibinfo{volume}{6}},
  \bibinfo{pages}{444} (\bibinfo{year}{2014}), ISSN \bibinfo{issn}{2073-8994},
  \urlprefix\url{http://www.mdpi.com/2073-8994/6/2/444}.

\bibitem[{\citenamefont{Chen et~al.}(2013)\citenamefont{Chen, Porada, Hooper,
  Klittnick, Shen, Tuchband, Korblova, Bedrov, Walba, Glaser
  et~al.}}]{Chen2013}
\bibinfo{author}{\bibfnamefont{D.}~\bibnamefont{Chen}},
  \bibinfo{author}{\bibfnamefont{J.~H.} \bibnamefont{Porada}},
  \bibinfo{author}{\bibfnamefont{J.~B.} \bibnamefont{Hooper}},
  \bibinfo{author}{\bibfnamefont{A.}~\bibnamefont{Klittnick}},
  \bibinfo{author}{\bibfnamefont{Y.}~\bibnamefont{Shen}},
  \bibinfo{author}{\bibfnamefont{M.~R.} \bibnamefont{Tuchband}},
  \bibinfo{author}{\bibfnamefont{E.}~\bibnamefont{Korblova}},
  \bibinfo{author}{\bibfnamefont{D.}~\bibnamefont{Bedrov}},
  \bibinfo{author}{\bibfnamefont{D.~M.} \bibnamefont{Walba}},
  \bibinfo{author}{\bibfnamefont{M.~A.} \bibnamefont{Glaser}},
  \bibnamefont{et~al.} (\bibinfo{year}{2013}), ISSN \bibinfo{issn}{0027-8424}.

\end{thebibliography}
\bibliographystyle{apsrev}

\end{document}